\documentclass[12pt,preprint]{aastex}

\usepackage{graphics,epsfig}
\usepackage{natbib}
\usepackage{lscape}

\def\spose#1{\hbox to 0pt{#1\hss}}
\def\lta{\mathrel{\spose{\lower 3pt\hbox{$\mathchar"218$}}
     \raise 2.0pt\hbox{$\mathchar"13C$}}}
\def\gta{\mathrel{\spose{\lower 3pt\hbox{$\mathchar"218$}}
     \raise 2.0pt\hbox{$\mathchar"13E$}}}

\def\figure#1#2 {\par{\narrower\noindent {\bf Fig. #1}
   \hskip 2mm #2\par}\bigskip\noindent}
\def\table#1#2 {\par{\narrower\noindent {\bf Tab. #1}
   \hskip 2mm #2\par}\bigskip\noindent}

\def\registered{{\ooalign{\hfil\raise .00ex\hbox{\scriptsize R}\hfil\crcr\mathhexbox20D}}}

\usepackage{amsmath}
\usepackage{accents}
\newlength{\dhatheight}

\shorttitle{The Case For Dwarf K Stars}

\shortauthors{Cuntz and Guinan}

\begin{document}


\title{About Exobiology: The Case for Dwarf K Stars
}

\author{M. Cuntz$^1$ and E. F. Guinan$^2$}

\affil{$^1$Department of Physics, University of Texas at Arlington, \\
Arlington, TX 76019, USA}
\email{cuntz@uta.edu}
\bigskip
\affil{$^2$Department of Astrophysics and Planetary Science, Villanova University, \\
Villanova, PA 19085, USA}
\email{edward.guinan@villanova.edu}

\begin{abstract}
One of the most fundamental topics of exobiology concerns the identification
of stars with environments consistent with life. Although it is believed that
most types of main-sequence stars might be able to support life, particularly
extremophiles, special requirements appear to be necessary for the development
and sustainability of advanced life forms. From our study, orange main-sequence
stars, ranging from spectral type late-G to mid-K (with a maximum at early-K),
are most promising. Our analysis considers a variety of aspects, including
(1) the frequency of the various types of stars, (2) the speed of stellar
evolution their lifetimes, (3) the size of the stellar climatological habitable
zones (CLI-HZs), (4) the strengths and persistence of their magnetic dynamo
generated X-ray--UV emissions, and (5) the frequency and severity of flares,
including superflares; both (4) and (5) greatly reduce the suitability of
red dwarfs to host life-bearing planets. The various phenomena show pronounced
dependencies on the stellar key parameters such as effective temperature and
mass, permitting the assessment of the astrobiological significance of
various types of stars. Thus, we developed a
``Habitable-Planetary-Real-Estate Parameter" (HabPREP) that provides a measure for stars
that are most suitable for planets with life. Early K stars are found to have
the highest HabPREP values, indicating they may be ``Goldilocks" stars for
life-hosting planets. Red dwarfs are numerous, having long lifetimes, but
their narrow CLI-HZs and hazards from magnetic activity make them less
suitable for hosting exolife. Moreover, we provide X-ray--FUV irradiances
for G0~V -- M5~V stars over a wide range of ages.
\end{abstract}


\clearpage


\section{Introduction}

Since the discovery of the first planet hosted by a Sun-like star was found
more than 20 years ago orbiting 51~Pegasi \citep{may95}, the search for life
in the Universe has received unprecedented attention.  It is now unequivocally
the most fundamental task of astrobiology, with significant ramifications
toward other fields such as astronomy, astrophysics, microbiology, planetary
and atmospheric science, and geodynamics.  Generally, the search for life
around stars, notably F--M type main-sequence stars, is mostly concentrated on
two pivotal aspects.  The first aspect deals with the identification of stars
suitable to offer sustained habitable environments, whereas the second aspect
focuses on planets through elucidating which planetary properties make them
(potentially) habitable.  Previously, significant progress has been made on
both aspects, including work by \cite{sca07}, \cite{lam09,lam13}, \cite{hor10},
\cite{rug13}, \cite{kas14}, among others.

Regarding the stellar aspect, a central theme is the identification of the
existence and stability of climatological habitable zones (CLI-HZs) aimed at
rocky planets of different sizes; see, e.g., \cite{kal12} for a updated
classification scheme.  Previous results pertaining to single stars have been
obtained by, e.g., \cite{kas93}, \cite{kop13,kop14} and for higher-order
(binary and multiple) systems by, e.g., \cite{cun14,cun15}.  Furthermore,
the impact of stellar evolution on the CLI-HZs has been explored as well
\citep[e.g.,][]{und03,rus13}.  These studies show that when stars evolve and
increase in luminosity with age, the CLI-HZs broaden and move outward.  Recent
work by \cite{kas14} considers updated CLI-HZ boundaries based on improved
planetary climate models and discusses remote life-detection criteria.  For
example, for the case of the Earth, the expected increase in the luminosity
of the Sun over the next $\sim$1--2 Gyr will result in the Earth no longer
comfortably residing within the solar CLI-HZ.

Other efforts concentrate on suitable stellar properties that are able to
provide favorable biospheric conditions for hosted planets.  This kind of work
explores the magnitude and temporal evolution of stellar magnetic-dynamo driven activity,
such as UV, EUV, and X-ray radiative emission \citep[e.g.,][]{gud97,gui02,rib05,tu15}
and (super-)flares \citep[e.g.,][]{haw03,mae12,gui16}.  \cite{gui03} pointed out
that for solar-like stars, the FUV and X-ray emissions of the young
(age $\sim$100 Myr) Sun could have been higher by 30--50 times and 100--500 times,
respectively, than today.  Generally, stellar activity is a crucial feature of
all late-type stars and most pronounced in M dwarfs; it subsides as stars age.
For solar-type and cooler stars (i.e., stars with outer convective zones), this
process is attributable to stellar angular evolution owing to internal changes
and to magnetized winds \citep[e.g.,][]{kep95,cha97}.

The aim of this work is to consider detailed knowledge about stars including
their relationships to planets to evaluate which types of stars constitute the
best and most likely candidates for the facilitation of long-term exobiology.
In Sect.~2, we discuss the amounts of habitable planetary real estate, as
determined by the sizes of the CLI-HZs and the relative frequency of the
various types of stars.  In Sect.~3, we consider stellar activity, i.e.,
high energetic stellar radiation, flares, superflares, and winds regarding their
relevance to the circumstellar environments. The overall assessment of habitability,
i.e., ``the big picture", is conveyed in Sect.~4.


\section{Habitable Planetary Real Estate Parameter (HabPREP)}

In the following, for the different types of main-sequence stars,
we focus on the size of the CLI-HZs and their relative frequency
to evaluate amounts of habitable planetary real estate.  Other
aspects such as the impact of magnetic-dynamo driven stellar activity
will be considered in Sect.~3.  Previously, \cite{kas93} utilized
1-D climate models, which were state-of-the-art at the time, to
compute the CLI-HZs for a domain of main-sequence stars with
spectral-types ranging from late F to early M. 
The basic premise of their work was the assumption of an Earth-type
planet with a CO$_2$/H$_2$O/N$_2$ atmosphere and, moreover, that
habitability requires the presence of liquid water on the planetary
surface.  This work was significantly improved through subsequent
studies, including the recent work by \cite{kop13,kop14}.  Their
approach encompassed numerous improvements, including the
consideration of revised H$_2$O and CO$_2$ absorption coefficients.

For example, the work by \cite{kop13,kop14} re-computes the
recent Venus / early Mars (RVEM) limits of the CLI-HZ
introduced by \cite{kas93}.  For a solar-like star, the range
of the RVEM extends from 0.75 to 1.77~au.  Other
limits regarding the CLI-HZs are based on the runaway greenhouse
effect (inner limit) and maximum greenhouse effect (outer limit);
in the following, these limits are used to signify the general
habitable zone (GHZ).  At the inner limit, the greenhouse phenomenon
is enhanced by water vapor, thus promoting surface warming, which
increases the atmospheric water vapor content, thus further raising
the planet's surface temperature.  Eventually, this will lead to
the rapid evaporation of all surface water.  For the outer limit,
it is assumed that a cloud-free CO$_2$ atmosphere shall still be able
to provide a surface temperature of $\sim$273~K (0$^{\circ}$~C).

For solar-like stars, and assuming an Earth-mass object,
\cite{kop13,kop14} identified the GHZ limits as approximately
0.95 and 1.68~au.  In contrast, \cite{kas93} identified them as
0.84 and 1.67~au, respectively.
Another limit of habitability is given as the moist greenhouse
limit, which for stars akin to the Sun has been updated to 0.99~au.
In the previous work by \cite{kas93} another limit was identified
given by the first CO$_2$ condensation obtained by the onset of
formation of CO$_2$ clouds at a temperature of 273~K, which has
not been supported by \cite{kop13,kop14}.  Nevertheless,
in the present work, the moist greenhouse limit and the limit
due to the first CO$_2$ condensation are used to define the
conservative habitable zone (CHZ), as motivated by a large array
of previous studies.  Note that in the work by \cite{kop13,kop14}
the CLI-HZ given by the RVEM limits is referred to as GHZ,
whereas the CLI-HZ between 0.95 and 1.68~au, as identified for
an Earth-mass planet, is referred to as CHZ.  This notation
is different from that of the present study, which however
closely follows previous conventions (see Table~1).

Results are given in Table~2.  An important aspect is that
compared to previously published versions of that figure,
updates have been made for stars of $T_{\rm eff} \lta 4500$
due to improved spectro-thermometry in consideration of
updated PHOENIX atmosphere models \citep{hus13,man13},
resulting in revised values for effective temperature, radius,
mass, and luminosity.  These results have been utilized for
the assessment and interpretation of {\it Kepler} field
planet--hosting candidates.  We also evaluated the stability
of the CLI-HZs for stars with spectral types from $\sim$F5~V
to $\sim$M5~V (see Table~2).
In this case, we explored when the inner limit of the CHZ / GHZ
overtakes the outer limit with the latter recorded at the beginning of
stellar main-sequence evolution.  This time of $t_{\rm ev}$,
usually referred to as timescale of the continuous habitable zone
(in reference to CHZ or GHZ), describing the region (or duration
of time) when a planet can be continuously habitable (i.e.,
able to maintain liquid water on its surface) is relatively
short for early and mid F-type stars, see also \cite{sat14}
for further results, but very prolonged for the more slowly
evolving, cooler, low-mass main-sequence K- and M-type stars.
For example, for K2~V stars, $t_{\rm ev}$ is identified as $\sim$22 and
$\sim$32 Gyr for the CHZ and GHZ, respectively, which is
more than a factor of 3 longer than for G2~V stars like
the Sun.  For low-mass M-type stars (i.e., red dwarfs;
$T_{\rm eff} \lta 3800$~K), $t_{\rm ev}$ is found to exceed 100~Gyr. 

Next, we focus on the relative frequency of stars, as obtained
by the initial mass function (IMF); see Figure~1 and 2. In this
regard, it is found that low-mass stars are
much more frequent than high-mass stars given by the shape of
the IMF \citep[e.g.,][]{kro01,kro02,cha03,cha05}.  The number of
stars strongly increases with decreasing stellar mass
(albeit uncertainties associated with the region of formation,
stellar metallicity, etc.), although the IMF shows some flattening
for stellar masses below $\sim$0.5~$M_\odot$ (i.e., mid and late
M-dwarfs), if displayed against ${\log}M$ with $M$ as stellar mass.
The IMF typically follows a broken power law, which is
an empirical function describing the distribution of initial masses
for a stellar population.  More than 75\% of stars are identified as
M-dwarfs.  Early work establishing the IMF has been given by, e.g.,
\cite{mue00}, \cite{luc00}, \cite{hil00}, and \cite{luh00}, which
focused on distinct stellar clusters including clusters located in
the Orion and Taurus constellation; some of this work also employed
the {\it Two-Micron All Sky Survey} (2MASS).  Several decades ago,
\cite{mil79} pointed out that the flatting of the IMF in the regime of
very low mass stars may be due to the mutual interactions between the 
fragments of interstellar clouds as well as their interactions with
the ambient gas rather than cloud fragmentation itself.  Later on,
this interpretation has been backed up by detailed numerical simulations
given by, e.g., \cite{cla08}, which also allowed to link the variations
of the initial IMF to the rate of star formation.  Note that the general
behavior of the IMF as obtained is also consistent with the number count
of stars in the solar neighborhood, i.e.,  the RECONS project\footnote{For
more information, and regular updates, visit http://www.recons.org.}
\citep{hen09,hen15}.

Forming the product between the sizes of the CLI-HZs and the relative
frequency of stars allows to describe the
``Habitable-Planetary-Real-Estate Parameter" (HabPREP) for the different types of
main-sequence stars\footnote{Strictly speaking, the definition of
HabPREP should also consider the timescale of stability for the continuous habitable
zone, $t_{\rm ev}$.  However, the omission of this step will not affect
our conclusions in a notable manner.  Stars between spectral type late-G
and M have highly stable CLI-HZ (see Table~2), whereas F-type stars as
well as early and mid-G stars have not.  Thus considering the impact
of $t_{\rm ev}$ regarding HabPREP would make the maximum between
stellar types G8~V and K2~V even more pronounced.}.
The results reveal a maximum for stars with
effective temperatures between about 4900~K and 5300~K
(i.e., K2~V --– G8~V stars) and a strong increase for M dwarfs of
effective temperatures less than about 4000~K.  The overall behavior
of HabPREP is determined by two separate trends.  First, the width
of the CLI-HZs steadily decreases as a function of decreasing
effective temperature or mass.  Second, the relative frequency of
stars according to, e.g., \cite{kro02} and \cite{cha03}, and first
increases, but then near 0.5~$M_\odot$ (with the exact value dependent
on the set of observational data and the particulars of the statistical
approach) levels off or starts to decrease (if displayed against
${\log}M$ with $M$ as stellar mass); both behaviors combined
result in a maximum for HabPREP as said (see Figure~1).
HabPREP is normalized to
unity for G2~V stars, following the IMF obtained by \cite{cha05},
based on the GHZ as choice for the CLI-HZ.  However, the general
behavior of the HabPREP function (e.g., position of its maximum)
shows little dependence on the type of CLI-HZ selected;
see Figure~2 for details.  As discussed in Sect. 3,
M dwarfs (even though preferred by HabPREP) will be ruled less
favorable for providing habitable
environments compared to orange dwarfs of spectral type late-G
to mid-K owing to their excessively high amounts of activity.


\section{Impact of High Energy X-ray--UV Radiation, (Super-)Flares and Winds}

Next we discuss the impact of high energy radiation, flares, and winds on the
prospect of habitability for HZ planets hosted by G- to M-type stars.
There is a number of earlier studies of solar-type stars on the influence
of dynamo-generated energetic radiation encompassing various wavelength
regimes, especially X-ray, extreme-UV (EUV) and far-UV (FUV) as well as
stars of different ages \citep[e.g.,][]{gud97,gui02,gui03,rib05}.
Previously, \cite{rib05} presented results from the ``Sun in Time" program
indicating that the coronal--transition region {X-ray}--EUV
($\sim$1--910 {\AA}) emissions of the young main-sequence Sun were
$\sim$100 (EUV) to $\sim$600 (X-ray) times stronger than those of the present Sun.
Similarly, the transition region and chromospheric FUV–-UV emissions of
the young Sun\footnote{Based on studies by, e.g., \cite{sch08} and \cite{mow12}
it has been found that the effective temperature of the young Sun was
$\sim$5600~K (G6--7~V).  This value is fairly close to the upper limit of
the range of stars regarded as favorable for supporting complex life forms
(see Sect.~4), although the emergence of sophisticated life on Earth occurred
at a considerably later stage when the solar temperature and luminosity had
increased.} were identified to be 20--60 and 10--20 times stronger, respectively,
than at present.

Even though Earth is a striking counterexample as it was able to survive
significant energetic radiation and wind flows from the early Sun owing
to magnetic protection \citep[e.g.,][]{gri04}, high levels of stellar
activity are typically considered a significant hindrance to the development
and sustainability of advanced exobiology.  Thus, we should also consider
the adverse effects that the active young Sun had on the two other initial
solar-HZ planets.  Because of high levels of magnetic activity of the early
Sun, Venus lost its original water inventory very early and now is a very hot,
dry, inhospitable planet \citep[e.g.,][]{kul06}.  Also, because of the Sun's
past high activity, coupled with the loss of its protective geomagnetic
field some 3.5~Gyr ago, Mars today is too cold and dry for complex life at
least on its surface \citep[e.g.,][]{fai10}.  Thus, it could be considered
very fortunate that the Earth's atmosphere, water inventories and life
survived, persisted, and evolved in spite of the harsh effects of the
active early Sun as well as the devastating effects of impacts of asteroids
and comets.  Therefore, in the view of Fermi's paradox, i.e., no signs or
signals of advanced life, see \cite{cho16} for recent discussions,
complex life could indeed be very rare.

In this study we mostly focus on stellar coronal X-ray and chromospheric
Ly-$\alpha$ fluxes of G0~V -– M5~V stars over a wide age range (see Figure~3).
Ly-$\alpha$ serves as an excellent indicator for FUV emission because this
emission line alone contributes 80--90\% of the total stellar FUV
(1150--1750~{\AA}) flux \citep[e.g.,][]{lin13}.
Emphasis is placed on young stars ($<1$ Gyr),
when fast rotation generates strong magnetic fields, which give rise to high
levels of X-ray and FUV emission.  A study of the X-ray and Ly-$\alpha$
properties of M-stars over a wide range of ages has recently been carried
out by \cite{gui16}.  In this study the X-ray and FUV Ly-$\alpha$ fluxes,
i.e., $f_{{\rm X}}$ and $f_{{\rm Ly}\alpha}$, were determined for a sample
of M0-5~V stars with ages from 0.1--11.5~Gyr for a nominal reference distance
of 1.0~au (and also for 0.17~au, the mid CLI-HZ for a $\sim$~M1~V star).  For
young M-stars (ages $<500$~Myr) the X-ray and $f_{{\rm Ly}\alpha}$ CLI-HZ
irradiances are both very high and comparable in strength.  As noted in this
study, the FUV Ly-alpha flux dominates the FUV flux, comprising 80--90\% of
the total 900--1800~{\AA} FUV flux and can thus be used to estimate the
total FUV irradiance. In the following, we extend the study of X-ray and
FUV (given by Ly-$\alpha$) irradiances of M-stars as a function of age,
to more massive and luminous G and K stars.

Note that the H~I Ly-$\alpha$ 1215.67~{\AA} emission line is by
far the strongest emission feature in the solar EUV--FUV ($\sim$100--1700{\AA}) 
spectrum of the Sun \citep{tia13}, and it is dominant
in the EUV--FUV regime of other solar-type (G stars) and cooler K and M
main-sequence stars as well.  Also, Ly-$\alpha$ emission is the main contributor
to the heating, ionization and photochemistry of the upper atmosphere of the Earth
as well as of many solar-system planets and moons \citep{hol84}; recent results have
also been obtained for planets in exosolar systems \citep[e.g.,][]{mig15}.  For Earth,
the Ly-$\alpha$ flux plays a major role in the photodissociation of important
molecules such as H$_2$O, CO$_2$, CH$_4$, O$_2$, and O$_3$ in planetary atmospheres.
Fortunately, reliable measures of Ly-$\alpha$ emission fluxes have meanwhile become
available, obtained with HST \citep[see, e.g.,][]{fra13,lin13}.  These stellar
Ly-$\alpha$ integrated fluxes have been reconstructed from HST-STIS and COS spectra
for a sample ($\sim$50) of main-sequence F5 -- M5 stars \citep{lin13,gui16}.

Similar to the M-star study of \cite{gui16}, we utilize the Ly-$\alpha$ and X-ray
flux measures of G and K stars obtained by \cite{lin13}. (Note that these authors
report those fluxes for a reference distance of 1.0~au from the star but they can
easily be transformed to any other distance following the inverse square law.)
These X-ray fluxes were supplemented by additional X-ray measures found in the
literature, previously obtained with ROSAT, XMM-Newton or Chandra. 
Ages were estimated from open cluster (such as Pleiades and Hyades) or moving group
memberships (such as the Ursae Majoris Moving Group), stellar rotation rates
(through employing rotation--age relations) or from memberships in wide binary
systems (such as Proxima Centauri)  in which one component has an isochronal
or astereoseismic age measures (e.g., $\alpha$~Cen).  X-ray fluxes are available
for a large number of stars from all-sky X-ray surveys like ROSAT as well as
from more recent X-ray observations by Chandra and XMM-Newton.  Mean milestone
age baseline X-ray flux--age calibrations of G and K stars were obtained
from ROSAT studies of Pleiades cluster (age $\sim$0.1~Gyr) and Hyades cluster
(age $\sim$0.65 Gyr) carried out by \cite{mic90} and \cite{ste95}, respectively.
These were supplemented by relations given by \cite{bas96}, \cite{per98}, and
\cite{mam08}.

Based on our exobiological perspective, we focus on planets at an Earth-equivalent
distance\footnote{Earth-equivalent distances mostly depend on the stellar luminosity
but the stellar effective temperature needs to be considered as well.  For example,
K and M-dwarfs have more emission in the near-IR compared to G-dwarfs, and a planet
can absorb more near-IR radiation from water-vapor around K and M-dwarfs making them
warmer at an Earth-equivalent distance if no effective temperature correction for
that distance is applied.  Previously, corrective formulae were given by
\cite{und03} and \cite{sel07} that are used by us, including recasting in response
to the studies of \cite{kop13,kop14}.},
also referred to as homeothermic distance (HTD), which for M0-6 dwarfs ranges between
$\sim$0.20 and $\sim$0.05~au; information for various stars is given in Figure~4
and Table~3. Our analysis also takes into account different $L_{\rm X}$ -- age
relations, which for G and M dwarfs have been obtained by \cite{gud07} and
\cite{gui16}, respectively. In addition, X-ray and FUV (Ly-$\alpha$) flux calibrations
were made for K-stars in similar manner as was done for M-stars. For intermediate
stellar types adequate interpolation is used.  Data for stellar activity at ages of
$\sim$5 Gyr are based on the Sun, 18~Sco, $\alpha$~Cen~A \& B, Proxima Cen, and
40~Eri~A \& C.  As shown in Figure~3 and 4, our analysis shows that for low-mass stars
(especially stars cooler than $\sim$M3~V), the amounts of planetary X-ray and FUV irradiance
at HTDs are drastically increased due higher {X-ray}--FUV emissions of young stars and
the closeness of the HTDs (i.e., $<0.2$~au) to their host stars.  Although in Table~3
we give the X-ray and FUV (Ly-$\alpha$) fluxes at the stellar HTDs, they can also be
applied to any other distance following the inverse square law.  Since the Ly-$\alpha$
emission flux contributes 80--90\% of the total FUV flux, multiplying the Ly-$\alpha$
flux by $\sim$1.15 should yield a good approximation for the total. 

As given in Table~3, and shown in Figure~3, the X-ray and Ly-$\alpha$ HTD fluxes
are comparable for young stars (age $<0.65$~Gyr) of all spectral types. At an age of
0.1~Gyr, the X-ray HTD fluxes are $\sim$1.5--2.0$\times$ higher than the corresponding
HTD Ly-$\alpha$ fluxes.  However, by an age of about 5 Gyr, the Ly-$\alpha$ HTD fluxes
are approximately 10--30 times larger than the corresponding X-ray HTD fluxes
across all spectral types.  Thus, we find that for older G--M stars, the
Ly-$\alpha$ HTD fluxes dominate the {X-ray}--FUV spectral region.  It is also seen
that the largest changes with stellar age occur for the X-ray HTD fluxes.
For example, for solar-type G2--8~V stars, the HTD X-ray flux is $\sim$400$\times$
higher for very young (0.1~Gyr) stars relative to older stars at ages of 5~Gyr.

The HTD X-ray and Ly-$\alpha$ fluxes are very high for stars cooler than
about M3~V compared to G and K stars of corresponding ages (see Figure~3).
For example, compared to a young (0.1~Gyr) G2~V star, the HTD X-ray flux
for an M4--5~V star ($T_{\rm eff} \sim 3200$~K) is over 45$\times$ higher.
This is primarily due to the very small values of HTDs for these
low-luminosity stars with HTD $<0.15$~au as well as their more efficient
magnetic dynamos due to their deep convective zones.  To illustrate this,
a HTD planet hosted by a solar-age ($\sim$5~Gyr) M5-6~V star (e.g.,
Proxima~Cen and Wolf~359) would receive over $\sim$400$\times$ and
$\sim$60$\times$ more X-ray and FUV radiation, respectively, than the
Earth presently receives from the Sun.  As shown in Table~2, at younger
ages ($\sim$0.1~Gyr) the HTD X-ray and FUV irradiances experienced by
planets hosted by M4--5~V red dwarfs are expected to be even higher.
For example, the X-ray and FUV HTD fluxes for these red dwarfs are inferred
as $\sim$55$\times$ and $\sim$6$\times$ higher, respectively, at 0.1~Gyr
compared to ages of $\sim$5~Gyr; furthermore, if compared to the irradiance
of today's Earth, they are elevated by even much higher factors, which are
$\sim$350 and $\sim$20,000 or more, respectively.

Another aspect of stellar activity pertains to flares and superflares.
Superflares are defined as having total energies of
$E > 10^{33}$~ergs \citep[see][]{mae12}.  Generally, it has been found
that flares are most intense and frequent in young stars and stars
of late spectral types, notably M-dwarfs.  Aside from the flare
properties (i.e., strength, spectral energy distribution, frequency,
stochasticity) effects on Earth-type planets in CLI-HZs are heavily
determined by the CLI-HZ's proximity, i.e., the very small values for
the HTDs for those stars.  Thus, detailed studies about the impact of
flares on possible exolife entail outcomes that are unfavorable or mixed
at best \citep[e.g.,][]{seg10,kas14}. Superflares as given by {\it Kepler}
data are found for stars of spectral types of G to M.
For example, \cite{can14} reported that from the more
than 100,000 stars included the study, 380 show superflares with a
total of 1690 such events.  With decreasing effective temperature,
an increase in the superflare rate is observed, which is consistent
to previous findings, and in alignment to dynamo theory.  They
also conclude that the resulting statistics of the dissipation
energy is similar to the observed flare statistics as a function
of the inverse Rossby number.  From the perspective of this study,
superflares are expected to be most significant for Earth-equivalent
planets in the consideration of their close proximity to the stars,
i.e., small HTDs.  The effects of flares on habitability of planets
is complex and beyond the scope of this paper; it will be considered
by us in a subsequent paper.

Dense stellar winds (as present in active young stars) are also
expected of having an adverse impact on circumstellar habitability
\citep{joh15}.  \cite{lam03} showed that the combination of high {X-ray}--UV
irradiance and strong (more dense) stellar winds expected from young
G, K, and M dwarfs can, via ion-pickup mechanisms, act synergically
to strip away atmospheres of close-in planets \citep[e.g.,][]{vid03}.
Unless a hosted CLI-HZ planet has a strong and persistent magnetic field,
amounting to a protective magnetosphere, there is a significant possibility
that the planet will lose most, if not all, of its atmosphere including its
water inventories \citep[e.g.,][]{gri04}.  In our solar system, the
early loss of water on Venus \citep{kul06} and $\sim$1~Gyr later on Mars
\citep[e.g.,][and references therein]{fai10} are best explained by
the lack of strong, sustained geomagnetic fields on these planets.
However, currently we are not in the position to include stellar
mass loss fluxes in this study largely because it is still uncertain
how the stellar wind properties change with spectral type and age.  For
example, the most recent wind measurement for $\kappa$~Ceti (G5~V),
a young ($\sim$0.7~Gyr) solar-type star, by \cite{don16} indicates
a mass loss rate of $\sim$50 times the present Sun.  This result agrees
very well with recent wind density estimates of solar-type stars with
age by \cite{air16a}.  Since stellar winds scale to magnetic fields and
activity \citep[e.g.,][]{sch03,pre05,cra11}, it can be assumed that winds are
most dense in young, magnetically active stars and planets at close-in HTDs
(as given for late-K and M dwarfs) are most affected.


\section{The Big Picture}

The aim of this work is a multi-facet attempt to identify types of stars
consistent with the durable existence of life, notwithstanding extremophiles,
for which according to terrestrial analyses \citep[e.g.,][]{rot01} large
windows of opportunities might exist.  General stellar aspects motivate us
to focus on the main-sequence rather than pre- or post-main-sequence scenarios,
which are typically of highly transitory nature.  Aspects of primary importance
include the frequency of the various types of stars, the size of the
stellar CLI-HZs, and the rapidness of stellar evolution for various types of
main-sequence stars.  Following previous work by, e.g., \cite{kop13,kop14}
that indicates the decrease in the size of the CLI-HZs with decreasing stellar
mass, luminosity and effective temperature --— irrespectively of the planetary
climatological criteria for defining the CLI-HZ limits --— tend to favor
higher-mass stars.  That is {F--G}-type stars have wide CLI-HZ limits whereas
lower mass mid-K and M-type stars have narrow CLI-HZs located close to the star.
Low-mass stars, however, are much more frequent than high-mass stars as given
by the shape of the IMF \citep[e.g.,][]{kro01,kro02,cha05}.
The total number of stars greatly
increases with decreasing stellar mass, although there is an onset of flattening
or modest decrease for masses below $\sim$0.5~$M_\odot$.  Hence, more than 75\% of stars are
identified as M-type dwarfs.  This result is also consistent with the number
count of stars in the solar neighborhood, i.e., the RECONS project
\citep{hen09,hen15}.

Based on the combined behaviors of the CLI-HZs and the IMF, we are able to conclude
that, from a statistical perspective, orange dwarf stars, ranging from spectral
type late-G to mid-K are most promising regarding long-term exobiology.  More
massive stars ($M \gta 1.3~M_\odot$), such as F-type stars, are known to rapidly
evolve away from the main-sequence \citep[e.g.,][]{mey93,mow12}, a significant
disadvantage for the evolution and sustainment of exolife, though limited
opportunities for circumstellar habitability may still exist
\citep[e.g.,][]{coc99,sat14}. Regarding M-type dwarfs, numerous studies
have been pursued about the prospects of providing habitable environments
\citep[e.g.,][]{seg05,lam07,tar07,sca07,lis07,kas14}.  Generally, these
studies convey multiple adverse aspects regarding supporting habitable environments,
including (but not limited to) intense high-energy radiative emissions, strong stellar
flares, the narrowness of the CLI-HZ, and possible geodynamic planetary insufficiencies
(e.g., lack of volatiles).

To date, there has been a large array of research targeting both flares and energetic
radiation for stars of different ages and spectral types.  Examples of flare studies
include work by, e.g., \cite{haw03}, \cite{rob05}, and \cite{dav12}.  Generally, it
has been found that flares are most intense and frequent in young stars and stars of late
spectral types, notably M-dwarfs \citep[e.g.,][]{fei07}.  Aside from the flare properties
(i.e., strength, spectral energy distribution, frequency, stochasticity) effects on
Earth-type planets in CLI-HZs are heavily determined by the CLI-HZ's proximity, i.e.,
the small values ($<0.2$~au) for the HTDs of those stars.  Thus, detailed studies about
the impact of flares on possible exolife entail outcomes that are unfavorable or mixed
at best \citep[e.g.,][]{seg10,kas14}.  Superflares revealed by {\it Kepler} data are found
for stars of spectral types of G to M, but again are expected to have most adverse
impacts for planets of M-dwarfs.  A large number of superflares have also
been detected for solar-type stars \citep[e.g.,][]{mae12,mae15,shi13,kat15}.
Surprisingly, superflares may have favorable ramifications for possible exolife
around G-type (and perhaps early K-type) stars \citep{air16b}, as they might trigger
the production of hydrogen cyanide (HCN), an essential molecule of prebiotic chemistry.

In the present work, we focus on the X-ray and FUV Ly-$\alpha$ irradiances for planets
in the CLI-HZ, notably located at the HTD for stars of different effective temperatures,
luminosities, and ages.  For most stars the X-ray fluxes dominate the EUV fluxes and
the Ly-$\alpha$ emission fluxes dominate in the FUV region and thus are both critical
(see Table 3 and Figure~3).
Ultimately, high levels of energetic radiation lead to significant planetary atmospheric
evaporation, as discussed by, e.g., \cite{lam03}, \cite{vid03}, and \cite{pen08}.
This behavior entails adverse consequences for the habitable environments of late-K and M
dwarfs (see Figure~4), which is in part also caused the very close proximity of the CLI-HZs.
Additionally, as argued by \cite{ray07}, there is a decreased probability of habitable planet
formation around those stars.  It has also been pointed out that accreting planets, if formed,
subsequently located in the CLI-HZ around stars cooler than K5 (including the full range of
M-dwarfs) are most likely subjected to runaway greenhouse processes, and thus may lose
substantial amounts of water initially delivered to them \citep{ram14,lug15,tia15}, thus
emerging as (almost certainly) uninhabitable dry planets (``desert worlds").  Particularly,
\cite{tia15} argued that Earth-mass planets with Earth-like water contents orbiting M-dwarfs
have a 10--100 times reduced likelihood than around G dwarfs.

The results for red dwarfs, as discussed, add further exobiological relevance
in support of late-G to mid-K orange dwarfs (i.e., $T_{\rm eff}$ between approximately
4900~K and 5300~K) as more likely hosts for planets supporting
complex life.  Another intriguing aspect that also tends to support our main conclusion
is the onset of tidal locking, which for planets located in the CLI-HZ (both CHZ and GHZ),
pertaining to a timescale of $\sim$4.5 Gyr, occurs for stars with effective temperatures
close to 4800 (${\pm}200$)~K (i.e., for $\sim$K3~V stars).  This indicates that for planets
hosted by stars hotter and more luminous, due to their more distant CLI-HZs, tidal locking
is relatively unlikely (although it may still occur at a later evolutionary time), whereas
for planets around cooler stars, tidal locking is expected to have happened --— even
though it should be pointed out that tidal locking by itself does not necessarily exclude
habitability \cite[e.g.,][]{bar08}, although its absence leads to more thermally balanced
planetary climates.  However, tidal-locking reduces the rotation period of the planet
and thus possibly reduces the planet's protective geomagnetic field.  Without a robust
magnetic field, ion-pick up mechanisms, combined with strong {X-ray}--UV radiation, strong winds,
and flares, could strip the planet of its atmosphere \citep[e.g.,][]{lam08}.  On the other hand,
\cite{dri15} showed that tidally-locked terrestrial planets could have their geomagnetic
fields sustained (or even enhanced) from the tidal heating of their iron-nickel cores.

Another reason why K-dwarfs are expected to provide habitable environments stems
from recent planetary geodynamic studies.  \cite{haq16} investigated the impact of
limit cycles on the width of stellar habitable zones.  Limit cycles mean that planets
positioned in the CLI-HZs cannot maintain stable, warm climates, but rather
should oscillate between long, globally glaciated states and shorter periods of
climatic warmth.  \cite{haq16} argue that such conditions, similar to those
experienced on Earth (``Snowball Earth") would be disadvantageous to the development
of complex life, including intelligent life.  Thus, limit cycles reduce the usable
extent of CLI-HZs as they compromise habitability for planets near the CLI-HZs' outer
rim.  The authors point out that for planets around K and M dwarfs, limit cycles should
not occur, thus allowing to foster habitable environments based on this criterion.
However, \cite{haq16} view M-dwarfs as less suitable for exolife in part based on the
same reasoning as conveyed in this study.

Clearly, the search for planets around different types of stars continues at an
unprecedented pace.  A fairly recent example includes the detection of a possible
Earth-mass planet by \cite{dum12} orbiting $\alpha$~Cen~B (K1~V), a member of
the closest stellar system to the Sun, albeit this planet orbits very close to
the star is not located within the CLI-HZ.  Furthermore, five possible super-Earth
planets have been reported to orbit the nearby G8.5~V star $\tau$~Ceti \citep{tuo13}.
Two of these large Earth-size planet candidates (i.e., $\tau$~Ceti~e and f) orbit near
the inner and outer boundaries of this old star's continuously habitable zone
\citep{pag15} and are
thus potentially habitable. Terrestrial-size planets have also been reported for
several other nearby orange dwarfs as well as from the {\it Kepler}
Mission\footnote{For further information, see Planetary Habitability Laboratory,
{\tt http://phl.upr.edu.}}.  Finding habitable planets around low-mass
K--M stars continues to be a challenging task due to the impacts of the stellar
radiative environments, intense plasma fluxes, and the narrowness of the CLI-HZs.
Additional planets are expected from the extended {\it Kepler} Mission (K2),
continuing and new high precision spectroscopic radial velocity studies, as well
as from the {\it Transiting Exoplanet Survey Satellite} (TESS) and Gaia.

However, it is noteworthy that recent statistical analyses of {\it Kepler} Mission exoplanet
data indicate that a significant fraction (i.e., 15\% -- 25\%) of red dwarfs is expected
to host terrestrial-size planets within their CLI-HZs \citep[e.g.,][]{bon13,dre13,dre15}.
This implies that tens of billions Earth-size planets in the Milky Way alone hosted
by M-dwarfs could have a chance of being potentially habitable, even though potential life
forms would most likely constitute extremophiles by terrestrial standards.  Possible
scenarios include that these potentially habitable planets could have escaped the
strong X-ray--UV radiation of their host stars by forming beyond the stellar CLI-HZs.
Thereafter, they could have migrated into the CLI-HZs when their host stars had become
older and less active.  Another possibility is that the planets could have lost their
original atmospheres and water inventories at a much later time (ages $>2$~Gyr),
outgassed and/or captured water and gases from collisions with icy bodies (i.e.,
exocomets) \citep[e.g.,][]{bos16}.  It is also possible that some planets could have
been protected by strong geomagnetic fields, shielding them from the expected strong
X-ray--UV radiation and winds from their host stars when those were young and
highly active \citep[e.g.,][]{kho07,erk13}.

In conclusion, our study indicates that the very high levels of X-ray and FUV radiation
experienced by close-in CLI-HZ planets of cool stars, especially red dwarfs, is expected
to have negative (detrimental) consequences on the development of life on such planets.
All GHZ planets hosted by G--K--M stars experience high levels of X-ray--UV irradiances
when the host stars are young ($<0.5$~Gyr), including early Earth.  However, as
discussed in this paper, the close-in GHZ planets hosted by late-K and M dwarfs
experience at least an order of magnitude higher levels ionizing radiation and far more
extended periods than the Earth.  To make matters worse, there is a greater likelihood
of large flares from young stars.  The levels of the resulting
ionizing X-ray--UV radiation and energetic plasmas from flares
could erode or eliminate the planet's atmosphere and water inventories, thus greatly
reducing the suitability of the planet for sustained life.  These effects as well as
other factors (including biological and chemical bottlenecks, asteroid bombardments,
among others), as recently discussed by \cite{cho16} and references therein, could
greatly diminish the likelihood of development of multicellular complex life.  The
lack of evidence of technologically advanced civilizations such as radio signals and
other modes of one-way communication are referred to as the Fermi's Paradox, which
continues to remain unresolved \citep[see][]{cho16}.  As discussed in this paper,
decisive impediments (roadblocks) to life could be attributable to high energy radiation,
strong stellar winds and flares experienced by planets during the first billion
years after formation.


\acknowledgments
This research is supported by the NSF and NASA through
grants NSF/RUI-1009903, HST-GO-13020.001-A and Chandra
Award GO2-13020X to Villanova University (E.~F.~G.).  We are very
grateful for this support.  Furthermore, it has been supported in part by NASA
through the American Astronomical Society's Small Research Grant Program
as well as the SETI Institute (M.~C.).
We also wish to acknowledge the availability of HST data through
the MAST website hosted by the Space Telescope Science Institute
and the Chandra X-ray data through the HEASARC website hosted
by the Astrophysics Science Division at NASA/GSFC and the High Energy
Astrophysics Division of the Smithsonian Astrophysics
Observatory (SAO).
Furthermore, we would like to thank an anonymous referee for her/his
useful suggestions allowing us to improve the manuscript.
This research made use of public databases hosted by SIMBAD,
and maintained by CDS, Strasbourg, France.
Moreover, we wish to acknowledge assistance by Zhaopeng Wang with
computer graphics.

Facilities: ROSAT, XMM-Newton, Chandra, HST (COS), {\it Kepler}



\clearpage


\begin{figure*} 
\centering
\begin{tabular}{c}
\epsfig{file=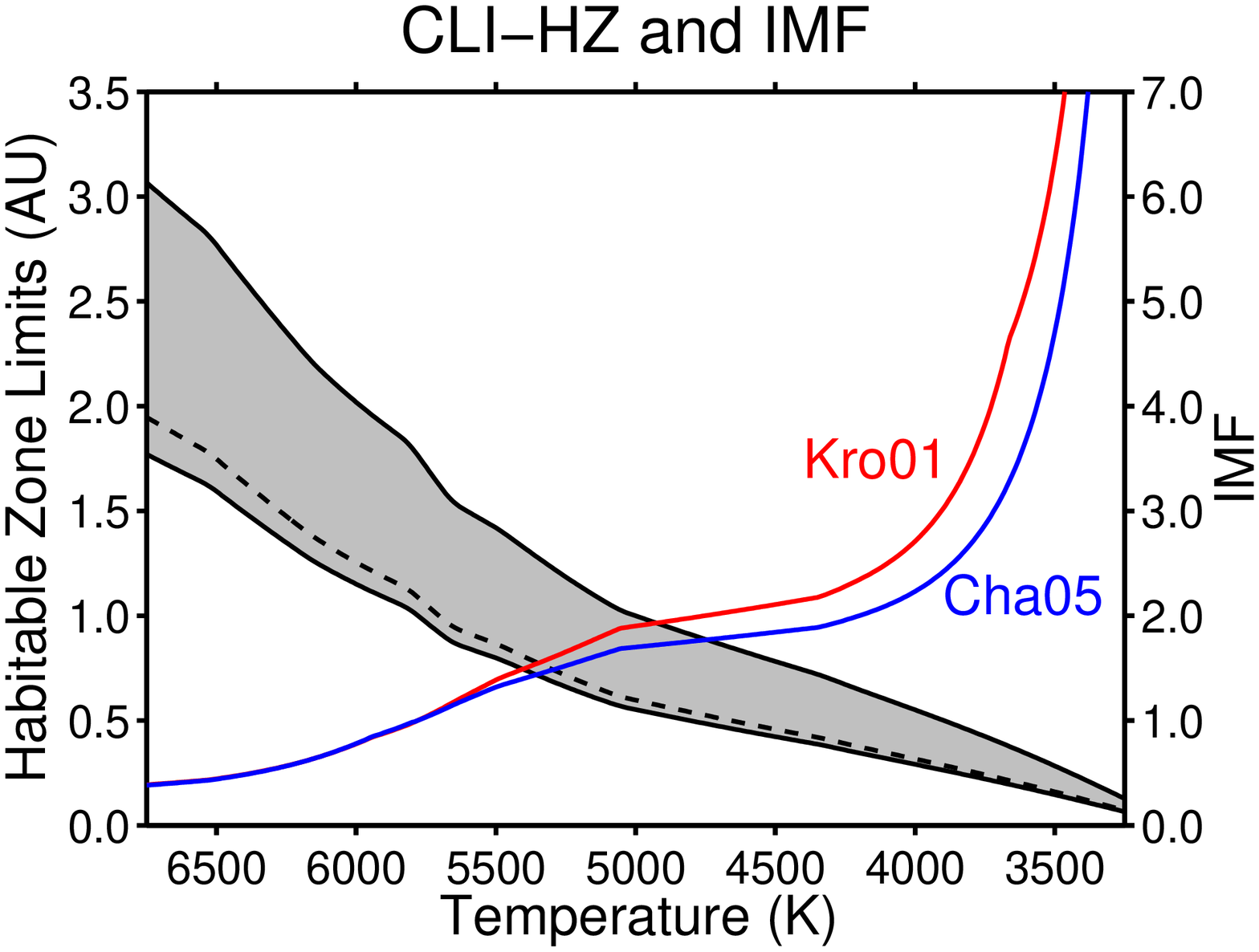,width=0.75\linewidth} \\
\end{tabular}
\caption{
Inner and outer limits of the CLI-HZ, represented by the GHZ (black lines)
with its extent depicted as grayish area.  The HTDs are depicted as well
(dashed line).  Additionally, we show the behavior of the IMF (normalized
to unity for 1~$M_\odot$) as given by \cite{kro01} (red line) and \cite{cha05}
(blue line).  Also note that IMF does not flatten near 0.5~$M_\odot$ if
displayed against the stellar effective temperature, though it does if
displayed against ${\log}M$ (with $M$ as stellar mass), the most customary
approach.
}
\end{figure*}

\clearpage


\begin{figure*} 
\centering
\begin{tabular}{c}
\epsfig{file=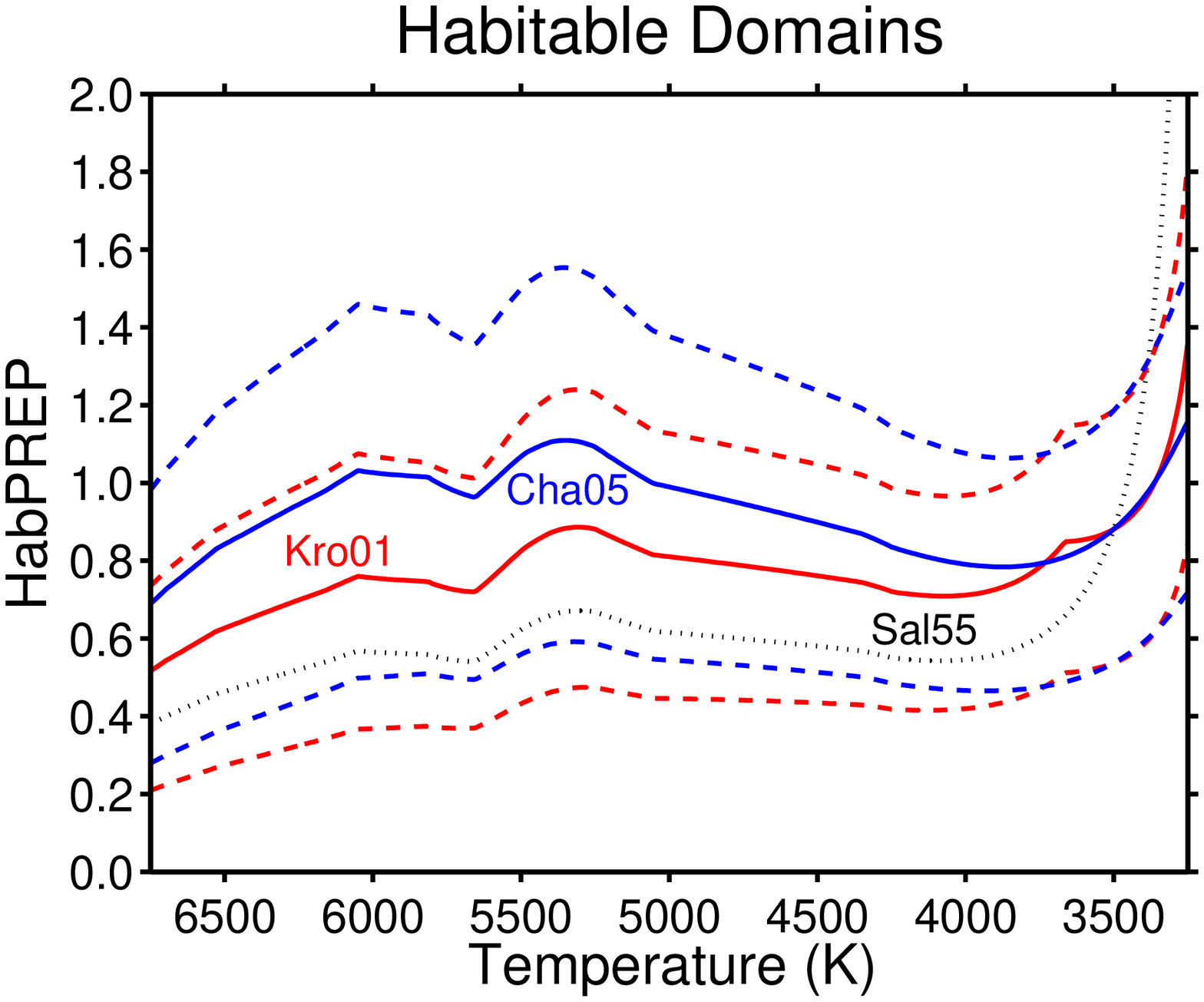,width=0.75\linewidth} \\
\end{tabular}
\caption{
HabPREP values depicted as function of the stellar effective temperature
with the IMF chosen after \cite{kro01} (red) and \cite{cha05} (blue).
Results are given regarding the GHZ (solid lines), RVEM (dashed lines, top),
and the CHZ (dashed lines, bottom).  For comparison and historic
reasons, the HabPREP values for the GHZ have also been given regarding the
IMF of \cite{sal55} (dotted line).  Note that the overall behavior of HabPREP
shows little dependence on the type of the CLI-HZ, i.e., RVEM, GHZ, or CHZ.
}
\end{figure*}

\clearpage


\begin{figure*} 
\centering
\begin{tabular}{c}
\epsfig{file=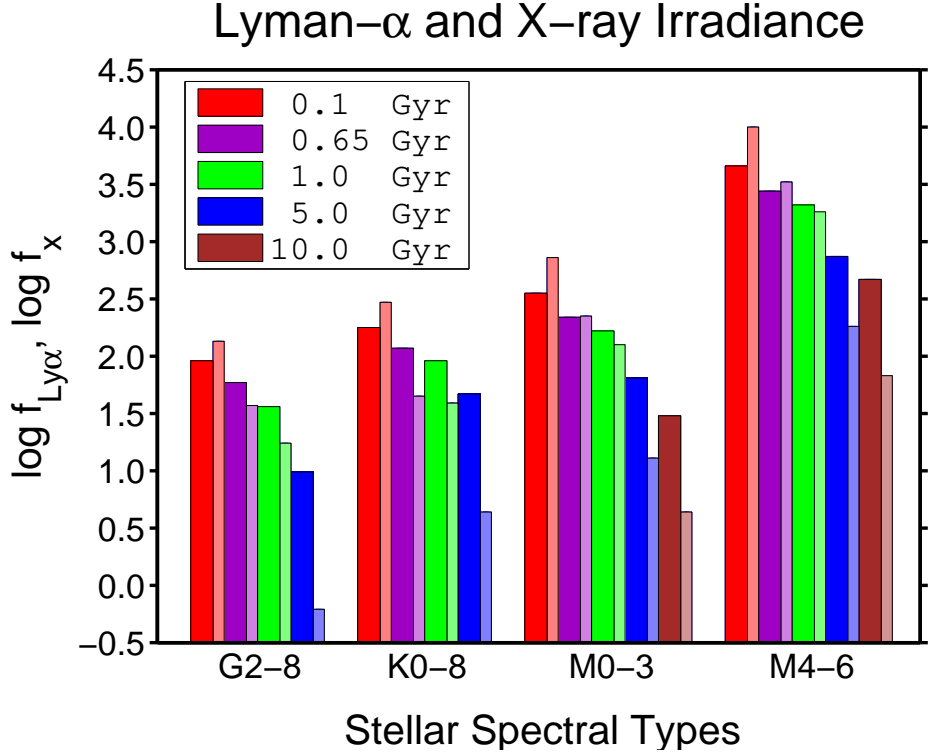,width=0.75\linewidth} \\
\end{tabular}
\caption{Lyman-$\alpha$ and X-ray irradiances (in ergs~{s$^{-1}$}{cm$^{-2}$})
for different sets of main-sequence stars regarding planets located at HTDs.
Stellar ages of about 0.1, 0.65, 1.0, 5.0, and 10~Gyr are represented by the colors
{\it red}, {\it purple}, {\it green}, {\it blue}, and {\it brown}, respectively.
Large-width bars indicate the results for Lyman-$\alpha$, whereas small-width
bars (with slightly lesser bright colors) convey the results for the X-ray
irradiances.  Note that the $y$-axis uses logarithmic units, allowing us to
display the drastic increase of both $f_{{\rm Ly}\alpha}$ and $f_{\rm X}$
between G dwarfs and late M dwarfs.
For the various G, K and M spectral type bins, the X-ray and FUV Ly-$\alpha$
HTD fluxes all increase with decreasing stellar age.  The increase of the HTD
X-ray flux from 0.1 to 5.0 Gyr is up to 500 times, whereas for the same age
range, the increase in FUV Ly-$\alpha$ HTD fluxes is much less (see also Table~3).
The latter is typically $\sim$5--10 times larger for young stars than for old
stars.  Moreover, the X-ray and FUV Ly-$\alpha$ HTD fluxes for stars of
similar ages from G2--8 dwarfs to M4--6 dwarfs increase over 100--500 times.
Thus HTD planets hosted by an M4--6 dwarfs (due to the very small HTDs) have
extremely high levels of X-ray and FUV irradiances, particularly when young.
}
\end{figure*}

\clearpage


\begin{figure*} 
\centering
\begin{tabular}{c}
\epsfig{file=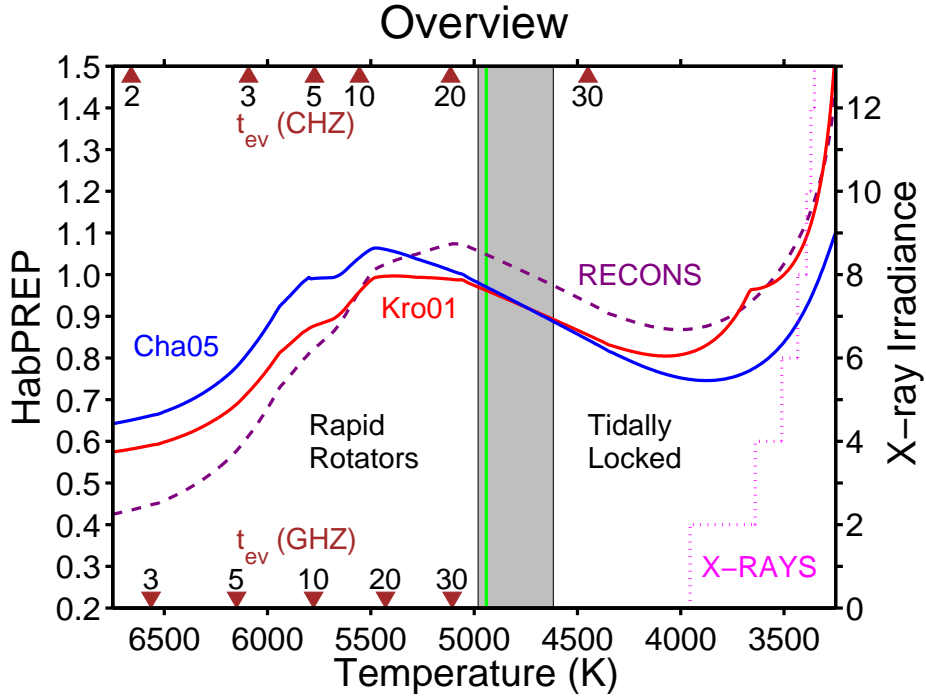,width=0.75\linewidth} \\
\end{tabular}
\caption{
Depiction of various functions and quantities, indicating the significance
(suitability) of late-G ($\sim$G8~V) to mid-K ($\sim$K5~V) type stars (i.e.,
``dwarf orange stars") for exobiology. The normalized habitable real estate,
defined on the left $y$-axis of the plot by the product of the
initial mass function and the HZ extent (IMF $\times$ HZ)
is displayed versus the stellar effective temperature.  As shown, the amount
of habitable planetary real estate, HabPREP, with the IMF given by \cite{kro01}
and \cite{cha05} as well as based on the results from the RECONS project
(see text), forms an aggregate maximum at the temperature range for
orange dwarfs.  We also convey the values of $t_{\rm ev}$ (in Gyr) for the
CHZ and GHZ, describing the timescales of stability for the continuous
habitable zone (with data beyond 30~Gyr disregarded).
For a tidal locking timescale of 4.5 Gyr, we show the domain for the onset of
tidal locking for planets located in the GHZs (grayish area).  The dividing
line (green) separates tidal locking and fast rotations for planets at HTDs.
Moreover, in the form of a histogram (magenta) defined on the right $y$-axis
of the plot, we show domains of possible exobiological exclusion owing to
high levels of X-ray irradiance.  We report distinct factors of
enhancements in $f_{{\rm X}}(650~{\rm Myr})$ for planets at HTDs
relative to a solar-type G2~V star of that age.
For example, the values for X-ray irradiance are found to be about 2$\times$,
5$\times$, and 10$\times$ higher for stars of $T_{\rm eff} \simeq 3950$~K
($\sim$K8~V), 3570~K ($\sim$M2~V), and 3390~K ($\sim$M3~V), respectively.
Factors larger than 100 are identified for stars of $T_{\rm eff} < 3085$~K
($\sim$M5~V).  For the latter, the CLI-HZ is
very close ($<0.05$~au) to the host star.
}
\end{figure*}

\clearpage


\begin{deluxetable}{lccccc}
\tablecaption{Habitable Zone Limits and Definitions}
\tablewidth{0pt}
\tablehead{
Description & \multicolumn{2}{c}{Models} & \multicolumn{3}{c}{Definitions} \\
\noalign{\smallskip}
\hline
\noalign{\smallskip}
...         &          Kas93 & Kop1314 & Kas93 & Kop1314 & This work           \\
\noalign{\smallskip}
}
\startdata
Recent Venus                 &  0.75   & 0.75  &  RVEM  &  optimistic   &  RVEM         \\
Runaway greenhouse effect    &  0.84   & 0.95  &  GHZ   &  conservative &  GHZ          \\
Moist greenhouse effect      &  0.95   & 0.99  &  CHZ   &  ...          &  CHZ$^{\ast}$ \\
First CO$_2$ condensation    &  1.37   & ...   &  CHZ   &  ...          &  CHZ$^{\ast}$ \\
Maximum greenhouse effect    &  1.67   & 1.68  &  GHZ   &  conservative &  GHZ          \\
Early Mars                   &  1.77   & 1.77  &  RVEM  &  optimistic   &  RVEM         \\
\enddata
\tablecomments{
Considering that the ``first CO$_2$ condensation" limit is not supported by the work
of \cite{kop13,kop14} reduces the relevance of the CHZ, hence labelled as ($^{\ast}$).
Nevertheless, we still convey this limit to allow comparisons with previous work.
\cite{kop13} use the terms ``optimistic" and ``conservative" limits; however, in
several previous studies those limits were identified as the GHZ and CHZ limits,
respectively, rather than the RVEM and GHZ limits.
}
\end{deluxetable}

\clearpage


\begin{deluxetable}{lcccccc}
\tablecaption{CLI-HZ Evolutionary Time Scales and Limits}
\tablewidth{0pt}
\tablehead{
 Sp. Type & $T_{\rm eff}$ & \multicolumn{2}{c}{$t_{\rm ev}$} & \multicolumn{3}{c}{CLI-HZ Limits} \\
\noalign{\smallskip}
\hline
\noalign{\smallskip}
 ...  & ...   &  CHZ    &  GHZ   &  CHZ   &  GHZ   & RVEM \\
 ...  & (K)   &  (Gyr)  &  (Gyr) &  (au)  &  (au)  & (au)       
}
\startdata
  F0   &  7100 &   1.4  &  2.1   &   2.32 -- 2.87  &   2.16 -- 3.72  &  1.70 -- 3.93   \\
  F2   &  6830 &   1.8  &  2.4   &   1.97 -- 2.51  &   1.85 -- 3.20  &  1.46 -- 3.37   \\
  F5   &  6530 &   2.3  &  3.1   &   1.72 -- 2.24  &   1.62 -- 2.82  &  1.28 -- 2.97   \\
  F8   &  6150 &   2.8  &  5.0   &   1.33 -- 1.78  &   1.26 -- 2.21  &  1.00 -- 2.33   \\
  G0   &  5940 &   3.7  &  7.5   &   1.17 -- 1.58  &   1.11 -- 1.96  &  0.88 -- 2.06   \\
  G2   &  5780 &   4.8  &  9.9   &   0.99 -- 1.36  &   0.95 -- 1.68  &  0.75 -- 1.77   \\
  G5   &  5670 &   7.8  & 13.9   &   0.91 -- 1.26  &   0.87 -- 1.55  &  0.69 -- 1.63   \\
  G8   &  5460 &  11.8  & 19.2   &   0.82 -- 1.14  &   0.79 -- 1.40  &  0.62 -- 1.48   \\
  K0   &  5250 &  16.0  & 24.8   &   0.69 -- 0.98  &   0.67 -- 1.20  &  0.53 -- 1.27   \\
  K2   &  5050 &  21.9  & 32.0   &   0.59 -- 0.84  &   0.57 -- 1.02  &  0.45 -- 1.08   \\
  K5   &  4410 &  30.4  & 41.4   &   0.39 -- 0.58  &   0.38 -- 0.70  &  0.30 -- 0.74   \\
  K8   &  4000 &  $>$50 &  $>$50 &   0.27 -- 0.41  &   0.26 -- 0.49  &  0.21 -- 0.52   \\
  M0   &  3800 & $>$100 & $>$100 &   0.22 -- 0.34  &   0.21 -- 0.41  &  0.17 -- 0.43   \\
\enddata
\end{deluxetable}

\clearpage


\begin{deluxetable}{lccccccccccc}
\tablecaption{Planetary Lyman-$\alpha$ and X-ray Irradiance for G, K and M Dwarfs}
\tablewidth{0pt}
\tablehead{
$T_{\rm eff}$ & HTD  & \multicolumn{2}{c}{0.1~Gyr}  &
                       \multicolumn{2}{c}{0.65~Gyr} &
                       \multicolumn{2}{c}{1.0~Gyr}  &
                       \multicolumn{2}{c}{5.0~Gyr}  &
                       \multicolumn{2}{c}{10~Gyr}   \\
 (K)    &   (au)     & \multicolumn{2}{c}{(ergs~{s$^{-1}$}{cm$^{-2}$})} &
                       \multicolumn{2}{c}{(ergs~{s$^{-1}$}{cm$^{-2}$})} &
                       \multicolumn{2}{c}{(ergs~{s$^{-1}$}{cm$^{-2}$})} &
                       \multicolumn{2}{c}{(ergs~{s$^{-1}$}{cm$^{-2}$})} &
                       \multicolumn{2}{c}{(ergs~{s$^{-1}$}{cm$^{-2}$})} \\
\noalign{\smallskip}
\hline
\noalign{\smallskip}
 ...           &  ... &  $f_{{\rm Ly}\alpha}$  & $f_{{\rm X}}$ &
                         $f_{{\rm Ly}\alpha}$  & $f_{{\rm X}}$ &
                         $f_{{\rm Ly}\alpha}$  & $f_{{\rm X}}$ &
                         $f_{{\rm Ly}\alpha}$  & $f_{{\rm X}}$ &
                         $f_{{\rm Ly}\alpha}$  & $f_{{\rm X}}$
}
\startdata
 5780    & $\equiv$~1  &    75.5  &   112  &   49.2 &   32.5 &   29.2 &   14.2  &    6.60 &   0.224  &  ...   &   ...  \\
 5500    &     0.866   &   101    &   149  &   65.6 &   43.3 &   38.9 &   18.9  &    8.80 &   0.299  &  ...   &   ...  \\
 5300    &     0.745   &    97    &   146  &   63.8 &   35.5 &   41.4 &   18.5  &   13.9  &   1.32   &  ...   &   ...  \\
 5000    &     0.597   &   101    &   155  &   67.2 &   27.4 &   48.9 &   19.7  &   23.6  &   1.80   &  ...   &   ...  \\
 4800    &     0.539   &   111    &   171  &   74.0 &   23.6 &   57.4 &   21.9  &   30.2  &   2.51   &  ...   &   ...  \\
 4600    &     0.433   &   172    &   265  &  115   &   36.6 &   89.0 &   34.0  &   46.8  &   3.89   &  ...   &   ...  \\
 4200    &     0.375   &   191    &   354  &  127   &   48.8 &   98.3 &   45.3  &   50.7  &   5.19   &  ...   &   ...  \\
 3900    &	   0.287   &   235    &   389  &  154   &   70.5 &  120   &   54.3  &   59.7  &   6.00   &  ...   &   ...  \\
 3600    &	   0.194   &   293    &   560  &  186   &  147   &  142   &   90.2  &   61.9  &   9.47   &  17.8  &   2.58 \\
 3400    &	   0.126   &   410    &   901  &  249   &  300   &  188   &  162    &   66.4  &  16.4    &  42.2  &   6.11 \\
 3200    &	   0.053   &  2318    &  5091  & 1404   & 1694   & 1063   &  917    &  375    &  92.6    & 239    &  34.5  \\
 3000    &     0.031   &  6777    & 14886  & 4107   & 4955   & 3107   & 2682    & 1097    & 271      & 697    & 101    \\
\enddata
\tablecomments{
The fluxes pertain to distances given by the HTDs.
}

\end{deluxetable}

\end{document}